\documentclass[3p, onecolumn]{elsarticle}

\usepackage[normalem]{ulem}
\usepackage{hyperref}
\usepackage{epsfig, bm, amsfonts, amssymb}
\usepackage[normalem]{ulem}
\RequirePackage{color}

 \definecolor{MyDarkGreen}{rgb}{0.02,0.60,0.06}

\begin{document}

\begin{frontmatter}

\title{Effective and asymptotic criticality of structurally disordered magnets}

\author[A,B]{Maxym Dudka}
\author[A,B,C]{Mariana Krasnytska}
\author[D,E]{Juan J. Ruiz-Lorenzo}
\author[A,B,F,G]{Yurij Holovatch}
\address[A]{~Institute for Condensed Matter Physics, National Academy of Sciences of Ukraine, 79011 Lviv, Ukraine}
\address[B]{~${\mathbb L}^4$ Collaboration \& Doctoral College for the Statistical Physics of Complex Systems, Leipzig-Lorraine-Lviv-Coventry, Europe}
\address[C]{~Laboratoire de Physique et Chimie Th\'eoriques, Universit\'e de Lorraine - CNRS, UMR 7019, Nancy, B.P. 70239,
54506 Vandoeuvre les Nancy, Francee}
\address[D]{~Departamento de F\'{\i}sica
  and Instituto de Computaci\'{o}n Cient\'{\i}fica Avanzada (ICCAEx),
  Universidad de Extremadura, 06071 Badajoz, Spain}
\address[E]{Instituto de Biocomputaci\'{o}n y F\'{\i}sica de Sistemas Complejos
  (BIFI), 50018 Zaragoza, Spain}
\address[F]{~Centre for Fluid and Complex Systems, Coventry University, Coventry, CV1 5FB, United Kingdom}
\address[G]{~Complexity Science Hub Vienna, 1080 Vienna, Austria}

\begin{abstract}
 Changes in magnetic critical behaviour of quenched structurally-disordered magnets
 are usually exemplified in experiments and in MC simulations by diluted
systems consisting of magnetic and non-magnetic components. By our study we aim to show, that
similar effects can be observed not only for diluted magnets with non-magnetic impurities,
but may be implemented, e.g., by presence of two (and more) chemically different magnetic
components as well. To this end, we consider a model of the structurally-disordered quenched
magnet where all lattice sites are occupied by Ising-like spins of different length $L$.
In such random spin length Ising model the length $L$ of each spin is a random  variable
governed by the distribution function $p(L)$. We show that this model belongs to the
universality class of the site-diluted Ising model. This proves that both models are
described by the same values of asymptotic critical exponents. However, their effective
critical behaviour differs. As a case study we consider a quenched mixture of two different
magnets, with values of elementary magnetic moments $L_1=1$ and $L_2=s$, and of concentration
$c$ and $1-c$, correspondingly. We apply field-theoretical
renormalization group approach to analyze  the renormalization group flow for different initial conditions, triggered
by $s$ and $c$, and
to calculate effective critical exponents further away from the fixed points of the renormalization
group transformation. We show how the effective exponents are governed by difference in
properties of the magnetic components.
\end{abstract}


\end{frontmatter}


\section{Introduction}\label{I}

Beneath the most intriguing questions of modern condensed matter physics one should certainly name a problem of
influence of structural disorder on criticality \cite{Order_books,Hertz,Dotsenko95,Folk03,Pelissetto02,Holovatch02}.
Taken that a regular lattice (`ideal') magnet manifests a
second order phase transition into magnetically-ordered state and a critical behaviour governed by some
scaling laws, will these universal laws be altered by structural disorder
induced into the system? Although this question addresses
properties of matter in a narrow region of  a phase diagram in
the vicinity of a phase transition point, the answer on it is
of a great importance both from the fundamental reasons
(description of criticality arising in different systems
ranging from high-energy physics to cosmogony) \cite{Order_books} as well
as due to applications of structurally-disordered materials
in modern technologies  \cite{refrigerants}.

In the context of this paper we concentrate on the case, when the disorder
in structure is implemented via quenched random local transition temperature \cite{Grinstein76}.
An archetypal example is given by diluted uniaxial magnets Fe$_x$Zn$_{1-x}$F$_2$, Mn$_x$Zn$_{1-x}$F$_2$ obtained as a crystalline mixture of two compounds when
a corresponding diluted alloy is prepared by a substitution of a non-magnetic
isomorph ZnF$_2$ for its antiferromagnetic counterpart (FeF$_2$ or MnF$_2$) \cite{Birgenau83,Belanger86,Mitchell86}.
Relaxation times of such systems are much larger than typical observation
times which allows one to perform their theoretical and numerical analysis
in terms of quenched diluted random-site Ising model, Fig. \ref{fig1}{\bf b}, see e.g., Refs. \cite{Folk03} and \cite{Kompaniets21}
for more recent references. Experimental data about critical behaviour
of quenched diluted Heisenberg magnets has been reviewed in Refs. \cite{Egami84,Kaul85,Dudka03}, and more recently in Refs.
\cite{Dinh17,Bouzaiene20,Tozri22,Jaballah22}.

For such systems, it is well established by now that even a weak structural disorder
can modify their critical behaviour. The qualitative answer about the changes in the universality
class of the 2nd order phase transition into magnetically ordered state is given by the
heuristic Harris criterion \cite{Harris74}. It predicts that if the
heat-capacity critical exponent $\alpha_{\rm pure}$ of the pure system is
positive, i.e., the heat capacity diverges at the critical point,
then a quenched disorder (e.g., dilution) causes changes in the universality class.
In particular, values of universal critical exponents change. Consecutively, if $\alpha_{\rm pure}<0$
the universality class is the same as for the
homogeneous system. In $d=3$ dimensions, it puts uniaxial  magnets
on a special place. Indeed,  only for Ising-like magnets with a
scalar (i.e., $m=1$-component) order parameter
the heat capacity diverges at the critical point,
$\alpha_{\rm pure}(m=1)>0$,  whereas it does not diverge for the
easy plane (XY) and Heisenberg magnets,  $\alpha_{\rm pure}(m=2), \alpha_{\rm pure}(m=3) < 0$,
correspondingly, see Table \ref{tab1}.
Numerous theoretical, experimental and numerical studies confirmed this observation
and, in particular, lead to determination of accurate values of the asymptotic
critical exponents of the 3d random Ising magnets in the new universality class \cite{Folk03,Kompaniets21}.
In Table \ref{tab1} we list values of the critical exponents $\nu$, $\alpha$, and $\gamma$
 for
the correlation length, magnetic susceptibility and heat capacity
of pure and quenched diluted 3d Ising model as obtained by resummation of the
perturbative field-theoretical renormalization group expansions. There, we also give
the values of the exponents for pure 3d XY and Heisenberg models.

\begin{table}[]
    \centering
    \begin{tabular}{|l|c|c|c|}
\hline Model & $\nu$ &
$\alpha$ & $\gamma$\\
\hline
$m=1$ \cite{Guida98}&  $0.6304(13)$ &   $0.109(4)$ & $1.2396(13)$   \\
$m=2$ \cite{Guida98} &  $0.6703(15)$   &   $-0.011(4)$ & $1.3169(20)$ \\
$m=3$ \cite{Guida98} &   $0.7073(35)$   &  $-0.122(10)$  &   $1.389(50)$ \\
 $m=1$, diluted \cite{Pelissetto00}&   $0.678(10)$  &    $-0.034(30)$& $1.336(20)$  \\
\hline
\end{tabular}
    \caption{Asymtotic critical exponents
of pure and quenched diluted 3d Ising model ($m=1$) as well as of the XY
($m=2$) and Heisenberg ($m=3$) models as obtained by resummation of the
perturbative field-theoretical renormalization group expansions.}
    \label{tab1}
\end{table}

The above results concern values of the asymptotic exponents.
By definition, an asymptotic exponent $x$ governs singular behaviour of the observable ${\cal O}$ directly at the critical point $T_c$:
\begin{equation}\label{1.1}
x\equiv - \lim_{\tau\to 0}\frac{\ln {\cal O}(\tau)}{\ln \tau}, \hspace{3em} \tau=|T-T_c|/T_c\, .
\end{equation}
However, an approach to $T_c$ is characterized by non-universal effective critical exponents,
which are introduced to describe the behavior of a quantity in a certain
temperature interval \cite{Kouvel64,Riedel74}. These are defined as:
\begin{equation}\label{1.2}
x_{\rm eff}(\tau)\equiv - \frac{\ln {\cal O}(\tau)}{\ln \tau}\, .
\end{equation}
They are  non-universal, change values with distance to the critical point and coincide with the asymptotic critical exponents only at (or close to) the critical point.
For structurally disordered systems, an effective critical behaviour is especially  rich and together with temperature
distance to the critical point may be governed by such parameters as inhomogeneities concentration, their distribution, etc.
Needless to say, that the effective non-universal critical behaviour of structurally-disordered magnets does not obey Harris criterion
in a sense that the dilution modifies effective critical exponents irrespective of the divergence in the heat capacity
of the corresponding pure magnet. In particular, numerous variations in scaling exponents of Heisenberg-like diluted
magnets have been observed experimentally and interpreted as being effective exponents, not yet in an asymptotic regime.
This concerns scaling of isothermal susceptibility \cite{Zarai16,Makni17,Hou19,Chebaane21}, spontaneous magnetisation \cite{Makni17,Hou19,Chebaane21}, magnetocaloric
coefficient \cite{Gebara21}, etc. The same is true
for the diluted uniaxial Ising-like magnets: although the theory predicts changes in their universality class and
hence new values of the asymptotic exponents (see Table \ref{tab1}), an effective exponents found in experiments
may differ essentially \cite{Perumal03}.

From the theoretical perspective, there are attempts of qualitative prediction and quantitative description of the
effective critical behavior for weakly diluted quenched Ising- \cite{Janssen95,Folk00,Calabrese04} and Heisenberg-like  \cite{Dudka03,Calabrese04}  magnets
within the field-theoretical renormalization group (RG) picture. On the one hand, this method is
capable to reproduce accurate values of the asymptotic exponents,  on the other hand, it can also take into account possible
violations of universality further away from the critical point. This is achieved by calculation of the RG functions further away
from the fixed points of the RG transformation. In this way, the RG flow equations mimic approach to criticality whereas
different initial conditions correspond to different non-universal factors, inherent to any specific magnet.

Our present paper continues aforementioned studies of effective critical behaviour of structurally disordered magnets,
however it also serves as an attempt to look on the phenomenon from a different perspective.
Indeed, the most common examples of structurally-disordered magnets with weak random-$T_c$-like disorder that were considered so far
in experiments and in MC simulations are exemplified by diluted systems consisting of magnetic and non-magnetic components.
In turn, changes in the critical exponents of structurally-disordered magnets are often attributed merely to the presence of a
non-magnetic component. By our study we aim to show, that a similar effect may be observed not only for diluted magnets with
non-magnetic impurities, but it is rather caused by presence of a quenched structural disorder which may be implemented,
e.g., by presence of two (and more) chemically different magnetic components as well.  To put it in a different way,
we aim to show that this is the structural disorder itself that causes changes in the critical behaviour, whenever
it has a form of randomly distributed quenched magnetic and non-magnetic constituents or the form of random mixture
of two different magnetic compounds.

The rest of the paper is organized as follows. In the next section \ref{II} we describe the random spin length Ising
model and sketch the derivation of its effective
Hamiltonian. In section \ref{III} we consider particular cases of this model. We show that the
effective Hamiltonian of the Ising model with randomly distributed elementary spins of two different lengths
 has the same symmetry as that of the site-diluted Ising model containing magnetic and non-magnetic sites. This proves
that both systems belong to the same universality class and hence are described by the same values of asymptotic
critical exponents. Moreover, we get the ratio of bare couplings of both effective Hamiltonians. This allows,
in particular, to discriminate between different effective critical behaviours generated by both Hamiltonians.
The last is analysed in details in Section \ref{IV}. There, we apply the field-theoretical RG approach to analyze
the RG flow for different initial conditions and to calculate effective critical exponents further away from the
fixed points of the RG transformation. Doing so we also discuss possible differences in effective critical behaviour
of two models under consideration. Conclusions and outlook are summarized in Section \ref{V}.

\section{\label{II}Functional representation for the free energy of the random spin length
Ising model}

In this chapter we introduce a model of structurally disordered magnet that will
be analysed henceforth.  Doing so we will get a functional representation of its
free energy and define corresponding effective Hamiltonian. The last is a key
object of analysis in the RG framework \cite{Amit,rgbooks}. Systems of different microscopic origin that
are described in the critical region by the same effective Hamiltonian share the
same universal features
of critical behaviour: it is said that they belong to the same universality class.

\begin{figure}
\begin{center}
\includegraphics[height=5.5cm]{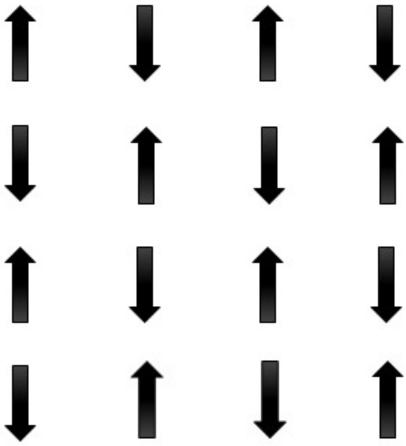}\hspace{15em}\includegraphics[height=5.5cm]{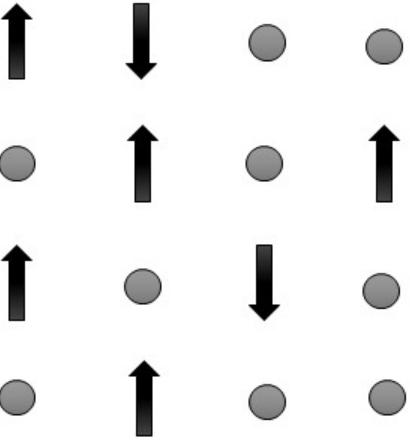}\\ \vspace{3ex}
\centerline{\bf (a) \hspace{25em} (b)} \vspace{3ex}
\includegraphics[height=5.5cm]{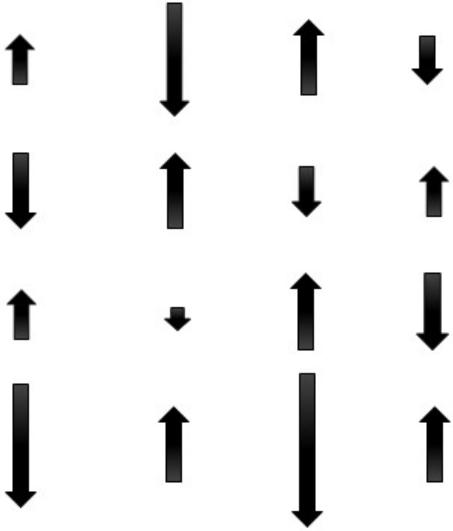}\hspace{15em}\includegraphics[height=5.5cm]{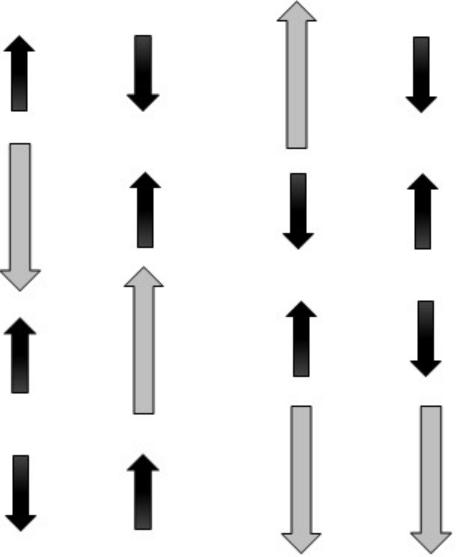}\\ \vspace{3ex}
\centerline{\bf (c) \hspace{25em} (d)} \vspace{3ex}
\end{center}
\caption{Lattice spin models discussed in this paper.
{\bf (a)} Ising model: magnetic atoms are described by pointing up and down
`spins' $S=\pm 1$ of equal length  located on sites of a $d$-dimensional lattice.
{\bf (b)} Quenched diluted Ising model: a part of sites in model (a) are occupied
by non-magnetic atoms (or are empty). Magnetic and non-magnetic
sites are randomly distributed and fixed in a given positions. Only magnetic atoms interact.
{\bf (c)} Random spin length Ising model: the length $L$ of a spin in model (a) is a random variable
governed by the distribution function $p(L)$, hence $S=\pm L$.  {\bf (d)}
A particular case of model (c) when the distribution function is given by Eq. (\ref{3.5}).
Note that model (b) can be also considered as a particular case of (c) with the distribution
function given by Eq. (\ref{3.2}).
\label{fig1}}
\end{figure}

It is convenient first to sketch a derivation of the effective Hamiltonian for the usual Ising
model and to use this derivation further as a blueprint to elaborate effective Hamiltonians incorporating structural disorder.
Let us consider a system of $N$  Ising spins $S_{\bf R}=\pm 1$ located
on ${\bf R}$ sites of a $d$ dimensional  hypercubic lattice that interact with the
Hamiltonian:
\begin{equation}
{\cal H} =  - \frac{1}{2}\sum_{{\bf R},{\bf R}'} J(|{\bf R}-{\bf R}'|)
S_{\bf R}S_{{\bf R}'}\, .
\label{2.1}
\end{equation}
Here and below sums (products) over ${\bf R}$ span all lattice sites.
In the case of the Ising model, the interaction $ J(R)>0$ is reduced to the nearest neighbours, but generally speaking one may consider any ferromagnetic short-range coupling. Symbolically, the model is depicted in Fig. \ref{fig1}{\bf a}.

Thermodynamics of the model can be analysed given its free energy
\begin{equation}
F = - \beta^{-1} \ln Z\,
\label{2.2}
\end{equation}
where $\beta=k_\mathrm{B} T$, $k_\mathrm{B}$ is the Boltzmann constant and the partition function reads
\begin{equation}
Z ={\rm Sp} \, {\rm e}^{-\beta {\cal H}}, \hspace{2em}
{\rm Sp} \, (\dots) = \prod_{\bf{R}} \sum_{S_{\bf{R}}=\pm 1}\, (\dots) \, .
\label{2.3}
\end{equation}
One way to take trace in Eq. (\ref{2.3}) is to represent the partition function in
a form of the functional integral to get rid of the product of two spins in the Hamiltonian (\ref{2.1}).
This can be achieved, e.g., by applying the Stratonovich-Hubbard transformation. To this end, making use of the translational invariance of the Hamiltonian (\ref{2.3})
one writes the partition function in the Fourier-space as a diagonal form:
\begin{equation}
 Z =  {\rm Sp} \, \prod_{{\bf k}} e^{\frac{\beta
\nu(k)}{2}{S}_{{\bf k}} {S}_{-{\bf k}}} \, , \label{2.4}
\end{equation}
here and below the wave vector ${\bf k}$ changes in the upper half of the corresponding Brillouin zone and
$\nu(k)$, ${S}_{{\bf k}}$ are the Fourier images of the interaction and spins:
\begin{eqnarray}
J(R) =  \frac{1}{ N} \sum_{{\bf k}} e^{i{\bf k}{\bf R}} \nu(k)\, , \hspace{2em}
\nu(k) =  \sum_{{\bf R}} e^{-i{\bf k}{\bf R}} J(R)\, , \label{2.5a} \\
S_{\bf R} =  \frac{1}{\sqrt{N}} \sum_{{\bf k}} e^{i{\bf k}{\bf R}} {S}_{{\bf k}}\, , \hspace{2em}
{S}_{{\bf k}} = \frac{1}{\sqrt{N}}  \sum_{{\bf R}} e^{-i{\bf k}{\bf R}} S_{\bf R}\, . \label{2.5b}
\end{eqnarray}
For the exponent in Eq. (\ref{2.4}),  the Stratonovich-Hubbard transformation is defined via the integral identity:
\begin{equation}
e^{\frac{\beta
\nu(k)}{2}{S}_{{\bf k}} {S}_{-{\bf k}}} = \sqrt{\frac{1}{2\pi\beta\nu(k)}}\int_{-\infty}^{\infty} {\rm d} {\phi}_{{\bf k}}
e^{- \frac{{\phi}_{{\bf k}} {\phi}_{-{\bf k}}}{2\beta\nu(k)} + {S}_{{\bf k}} {\phi}_{-{\bf k}}}
\label{2.6} \, ,    \hspace{2em} {\rm Re}\, \nu(k)>0\, ,
\end{equation}
with the functional integral over field ${\phi}_{{\bf k}}$ in the r.h.s.
This transformation allows
to reconsider a system of spins $S_{\bf R}$ interacting via the two-body
potentials $J(R)$  as a system of independent particles interacting with a fluctuating field ${\phi}_{{\bf k}}$. Further steps
of representing partition function of the $d$-dimensional Ising model in the form of a functional integral
can be sketched as:
\begin{equation}
Z \sim  \int ({\rm d} {\phi}) {\rm e}^{-\sum_{\bf k}\frac{1}{2\beta \nu(k)} {\phi}_{{\bf k}}
{\phi}_{{\bf -k}}} {\rm Sp} \ {\rm e}^{\sum_{\bf R}{S}_{{\bf R}}  {\phi}_{{\bf R}}}\sim
  \int ({\rm d} {\phi}) {\rm e}^{-\sum_{\bf k}\frac{1}{2\beta \nu(k)} {\phi}_{{\bf k}}
{\phi}_{{\bf -k}}} \prod_{\bf R} \cosh ({\phi}_{{\bf R}}) \sim \, \int ({\rm d} {\phi}) {\rm e}^{-{\cal H}_{\rm eff}}\, .
\label{2.7}
\end{equation}
Here and below we omit prefactors irrelevant for the subsequent analysis, ${\phi}_{\bf{R}}$ is the Fourier
image of ${\phi}_{{\bf k}}$ and the functional integration means:
\begin{equation}
\int ({\rm d} {\phi}) = \prod_{\bf{R}} \int_{-\infty}^{\infty} {\rm d} {\phi}_{\bf{R}}\, .
\label{2.8}
\end{equation}

Since the first expression in Eq.  (\ref{2.7}) contains only a linear in  ${S}_{{\bf R}}$ term, the trace merely reduces
 to substituting there the value ${S}_{{\bf R}} = \pm 1$ and leads to the product of $\cosh ({\phi}_{{\bf R}})$ in the second expression.
 In turn, in the critical region this function is approximated by an expansion up to the fourth order terms leading to
 the following effective Hamiltonian:
\begin{equation}
{\cal H}_{\rm eff} = \frac{1}{2}\sum_{\bf k}\big (\frac{1}{\beta \nu(k)}-u_2 \big ) {\phi}_{{\bf k}}
{\phi}_{{\bf -k}} + \frac{u_4}{4!} \sum_{\bf R} \phi_{\bf R}^4\, .
\label{2.9}
\end{equation}
Coefficients $u_{2l}$ readily follow from the expansion of $\ln\cosh(x)$ in Eq. (\ref{2.7}): $u_2/2=1/2$, $u_4/4!=1/12$.
In the RG analysis of critical behaviour the short wave-length (small $k$) contributions to $\nu(k)$ are considered,
the Gaussian term attains the form $\sim(k^2 + m^2) {\phi}_{{\bf k}}
{\phi}_{{\bf -k}}$, resulting in the familiar effective Hamiltonian (\ref{2.9}) of the
scalar $\phi^4$ model \cite{Amit,rgbooks}.

An impact of structural disorder on the critical behaviour of models (\ref{2.1}), (\ref{2.9}) is usually
considered within the paradigm of weak quenched dilution by a non-magnetic component, as shown in Fig. \ref{fig1}{\bf b}.
The resulting critical behaviour in this case is governed by the effective Hamiltonian containing two
 $\phi^4$ terms of different symmetry \cite{Folk03,Pelissetto02,Holovatch02}. We will return to this effective Hamiltonian later, see
 subsection \ref{IIa} below. Let us first consider the situation when structural disorder is implemented
by randomness in distribution of magnetic atoms of different species (different elementary magnetic moments).
The mentioned above weak quenched dilution by a non-magnetic component will follow as one of its limiting
cases. To this end, let us consider a set of spins that can point only up and down but are of different length,
as shown in Fig. \ref{fig1}{\bf c}. In such random spin length Ising model the length $L$ of each spin is a random
variable governed by the distribution function $p(L)$. The Hamiltonian reads \cite{Krasnytska20,Krasnytska21}:
\begin{equation}
{\cal H} =  - \frac{1}{2}\sum_{{\bf R},{\bf R}'} J(|{\bf R}-{\bf R}'|)
{\mathbb S}_{\bf R}{\mathbb S}_{{\bf R}'}\, , \hspace{2em}  {\mathbb S}_{\bf R} = \pm L_{\bf R} .
\label{2.10}
\end{equation}
where $ L_{\bf R}$ are i.i.d. random variables:
\begin{equation}
 \label{2.11}
P(\{L\})=\prod_{\bf R} p(L_{\bf R})\,
\end{equation}
and the rest of notations are the same as in Eq. (\ref{2.1}).
The Hamiltonian (\ref{2.10}) can be conveniently rewritten in terms of `usual' Ising
spin variables as:
\begin{equation}  \label{2.12}
{\cal H} =  - \frac{1}{2}\sum_{{\bf R},{\bf R}'} J(|{\bf R}-{\bf R}'|)
{S}_{\bf R}{S}_{{\bf R}'}{L}_{\bf R}{L}_{{\bf R}'}\, , \hspace{2em}  {S}_{\bf R} = \pm 1 .
\end{equation}
For the quenched disorder we are interested in, the spin lengths ${L}_{\bf R}$ are randomly distributed
and fixed in a certain configuration $\{L\}$. The partition function is configuration-dependent and physical observables are obtained from the
configurational average of the free energy \cite{Brout59}:
\begin{equation}\label{2.13}
 F=-\beta^{-1} \langle \ln Z (\{L\}) \rangle_{\{L\}}, \hspace{1em}
\langle \dots \rangle_{\{L\}} = \prod_{\bf R} \sum_{L_{\bf R}}p(L_{\bf R})(\dots)\, ,
\end{equation}
the sum over  $L_{\bf R}$ in (\ref{2.13}) spans all values of $L_{\bf R}$.
It is a usual practice to avoid averaging of the logarithm in Eq. (\ref{2.13}) making use of the replica trick \cite{Dotsenko01}
\begin{equation} \label{2.14}
 \ln Z (\{L\}) = \lim_{n\to 0} \frac{\big (Z (\{L\})\big )^n -1}{n}\,.
\end{equation}
An expression for the $n$th power of the configuration-dependent partition function reads:
\begin{eqnarray}\label{2.15}
\big ( Z(\{L \}) \big )^n &=& \prod_{\alpha=1}^n {\rm Sp}_{S^{\alpha}} {\rm e}^{ \frac{\beta}{2}\sum_{{\bf R},{\bf R}'} J(|{\bf R}-{\bf R}'|)
{S}_{\bf R}^\alpha{S}_{{\bf R}'}^\alpha{L}_{\bf R}{L}_{{\bf R}'}} \\ \nonumber
&\sim&  \prod_{\alpha=1}^n  \int ({\rm d} {\phi}^\alpha) {\rm e}^{-\sum_{\bf k}\frac{1}{2\beta \nu(k)} {\phi}^\alpha_{{\bf k}}
{\phi}^\alpha_{{\bf -k}}} \prod_{\bf R} \cosh ({\phi}^\alpha_{{\bf R}}L_{\bf R})\, .
\end{eqnarray}
Here, as in Eq. (\ref{2.7}), we have omitted
the pre-factor irrelevant for the subsequent analysis and made use of the Stratonovich-Hubbard transformation,
the field $\phi^\alpha$ holds now the replica index $\alpha$.
Now the last term in Eq. (\ref{2.15}) is to be averaged with respect to $\{L\}$:
\begin{eqnarray}\nonumber
\langle \big ( Z(\{L \}) \big )^n  \rangle_{\{L\}} &\sim& \prod_{\alpha=1}^n  \int ({\rm d} {\phi}^\alpha) {\rm e}^{-\sum_{\bf k}\frac{1}{2\beta \nu(k)} {\phi}^\alpha_{{\bf k}}
{\phi}^\alpha_{{\bf -k}}} \prod_{\bf R}  \langle \cosh ({\phi}^\alpha_{{\bf R}}L_{\bf R}) \rangle_{\{L\}}
\\ \label{2.16} &\sim&  \int ({\rm d} {\phi}) {\rm e}^{-{\cal H}_{\rm eff}}\, ,
\end{eqnarray}
with
\begin{equation}
\int ({\rm d} {\phi}) = \prod_{\alpha=1}^n \prod_{\bf{R}} \int_{-\infty}^{\infty} {\rm d} {\phi}^\alpha_{\bf{R}}\, ,
\end{equation}
leading to the effective Hamiltonian with two $\phi^4$ terms of different symmetry:
\begin{eqnarray}\nonumber
{\cal H}_{\rm eff} &=& \frac{1}{2}\sum_{\alpha=1}^n\sum_{\bf k}\big (\frac{1}{\beta \nu(k)}-u_2\langle L^2 \rangle \big ) {\phi}^\alpha_{{\bf k}}
{\phi}^\alpha_{{\bf -k}} + \frac{u_4}{4!}\langle L^4 \rangle \sum_{\alpha=1}^n \sum_{\bf R} (\phi^\alpha_{\bf R})^4\,  \\
\label{2.17} &-& \frac{u_2^2}{8}(\langle L^4 \rangle - \langle L^2 \rangle^2)  \sum_{\alpha,\beta=1}^n \sum_{\bf R} (\phi^\alpha_{\bf R})^2
(\phi^\beta_{\bf R})^2 \, .
\end{eqnarray}
Here, the coefficients $u_{2l}$ are the  same as those obtained above for the
effective Hamiltonian of the usual Ising model, cf. Eq. (\ref{2.9}), and the averaging
is performed with a local spin length distribution function $p(L)$:
\begin{equation}\label{2.18}
    \langle \dots \rangle = \sum_L p(L)(\dots)\, .
\end{equation}
To analyze thermodynamics of the random spin length Ising model
the replica limit $n\to 0$ is to be implemented.
Expression (\ref{2.17}) will be further considered in the next section for
different distributions of the random variables $\{ L \}$.

\section{\label{III}Case studies}
\subsection{Ising model with non-magnetic impurities\label{IIa}}
Let us consider first the familiar case of the diluted Ising model, when a part of lattice
sites  of concentration $c$  are occupied by magnetic atoms, the other $1-c$ sites
are empty, or occupied by non-magnetic atoms \cite{Stinchcombe}, i.e., the random variable $L$ attains
two values
\begin{equation}
L =  \left \{ \begin{array}{ll}
1, & \quad \mbox{with prob. $c$},
\\
0,
&
\quad \mbox{with prob. $(1-c)$}\, ,
\end{array}
\right.
\label{3.1}
\end{equation}
which corresponds to the following distribution function
in Eq. (\ref{2.18}):
\begin{equation}
p(L)= c \delta (L-1) +
(1-c) \delta (L)\, ,
\label{3.2}
\end{equation}
with $\delta(x)=\delta_{x,o}$ being Kronecker delta.
The diluted Ising model is symbolically shown in Fig. \ref{fig1}{\bf b}.
Calculating moments of the distribution (\ref{3.2}):
$$
\langle L^k \rangle = \langle L \rangle=c\, ,
$$
$$
\langle L^4 \rangle - \langle L^2 \rangle^2 = c - c^2=c(1-c)\, ,
$$
and substituting them into (\ref{2.18}) we get the following
expression for the effective Hamiltonian of the diluted Ising model:
\begin{equation}\label{3.3}
{\cal H}_{\rm eff} = \frac{1}{2}\sum_{\alpha=1}^n\sum_{\bf k}\big (\frac{1}{\beta \nu(k)}- c \big ) {\phi}^\alpha_{{\bf k}}
{\phi}^\alpha_{{\bf -k}} +
\frac{c}{12} \sum_{\alpha=1}^n\sum_{\bf R} (\phi^\alpha_{\bf R})^4\, -
\frac{c(1-c)}{8} \sum_{\alpha=1}^n \sum_{\beta=1}^n\sum_{\bf R}  (\phi^\alpha_{\bf R})^2 (\phi^\beta_{\bf R})^2 \, .
\end{equation}
\subsection{Ising model with spins of two different lengths\label{IIb}}

Now let us consider the case, when all lattice sites are occupied by spins,
however the length of some spins is fixed to 1, the others being of fixed length $s$, see Fig. \ref{fig1}{\bf d}.
Concentration of spins of length 1 is $c$ and concentration of spins
of length $s$ is $1-c$:
\begin{equation}
L
= \left \{ \begin{array}{ll}
1, & \quad \mbox{with prob. $c$},
\\
s,
&
\quad \mbox{with prob. $(1-c)$}.
\end{array}
\right.
\label{3.4}
\end{equation}
 When spins of both length are randomly distributed,
this corresponds to the following two-parameter distribution function
\begin{equation}
p(L)= c \delta (L-1) +
(1-c) \delta (L-s)\, .
\label{3.5}
\end{equation}
The moments of the random variable $L$ with distribution (\ref{3.5})
readily follow:
$$
\langle L^k \rangle = c + (1-c)s^k\, ,
$$
$$
\langle L^4 \rangle - \langle L^2 \rangle^2 = c(1-c)(1-s^2)^2\, .
$$
Substituting these expressions into (\ref{2.17}) we get the following
effective Hamiltonian:
\begin{eqnarray}\nonumber
{\cal H}_{\rm eff} &=& \frac{1}{2}\sum_{\alpha=1}^n\sum_{\bf k}\big (\frac{1}{\beta \nu(k)}- c \big ) {\phi}^\alpha_{{\bf k}}
{\phi}^\alpha_{{\bf -k}} +
\frac{c+(1-c)s^4}{12} \sum_{\alpha=1}^n \sum_{\bf R} (\phi^\alpha_{\bf R})^4\, \\ &-& \label{3.6}
\frac{c(1-c)(1-s^2)^2}{8}\sum_{\alpha=1}^n\sum_{\beta=1}^n\sum_{\bf R}(\phi^\alpha_{\bf R})^2(\phi^\beta_{\bf R})^2 \, .
\end{eqnarray}

The crucial feature of the effective Hamiltonians (\ref{3.3}) and (\ref{3.6}) is that they both possess $\phi^4$ terms
of the same symmetry. This means, that being treated by the RG approach they both give origin to the
same picture of the fixed points. In turn, this will result in the same universal critical behaviour, provided the stable
fixed point is reachable from the initial conditions. Returning back to the lattice models that where used to derive
the effective Hamiltonians, see Figs. \ref{fig1}{\bf b} and \ref{fig1}{\bf d}  this conclusion means that uniaxial magnets with quenched
nonmagnetic impurities and quenched alloys of uniaxial magnets that differ in elementary  magnetic moments belong to the same
universality class.  However, due to the difference in the numerical values of the coefficients in (\ref{3.3}) and (\ref{3.6}), the effective
critical behaviour may differ, as we will further outline below. One notices also that in the limiting case $s=0$ ($1-c$
spins are of `zero length', i.e., they correspond to non-magnetic sites) the  effective Hamiltonian (\ref{3.6} reduces to
Eq. (\ref{3.3}). Furthermore, at $s=1$ (i.e., all spins are of the same length) the last term
in Eq. (\ref{3.6}) vanishes and the effective Hamiltonian factorizes to $n$ effective Hamiltonians of the usual Ising model (\ref{2.9}).

\subsection{Comparison of cases \ref{IIa} and \ref{IIb}\label{IIc}}

For  further analysis it is convenient to introduce a ratio of the coefficients at two $\phi^4$ terms in Eq. (\ref{3.6}):
\begin{equation}\label{3.7}
 r(c,s)= -\frac{3}{2} \frac{c(1-c)(1-s^2)^2}{c+ (1-c)s^4}\, .
 \end{equation}
As discussed at the end of the former subsection, at $s=0$ the effective Hamiltonian (\ref{3.6}) corresponds to that
of the Ising model with non-magnetic impurities and hence $r(c,s=0)=-\frac{3}{2}(1-c)$, the result one obtains also dividing
the coefficients at the  $\phi^4$ terms of Eq. (\ref{3.3}). In turn,  $r(c,s=1)=0$, as corresponds to the usual Ising model. Another limiting case when the lengths of two species of spins essentially differ is $r(c,s=\infty)=-\frac{3}{2}c$.

\begin{figure}[th]
    \begin{center}
        \includegraphics[width=0.4\paperwidth]{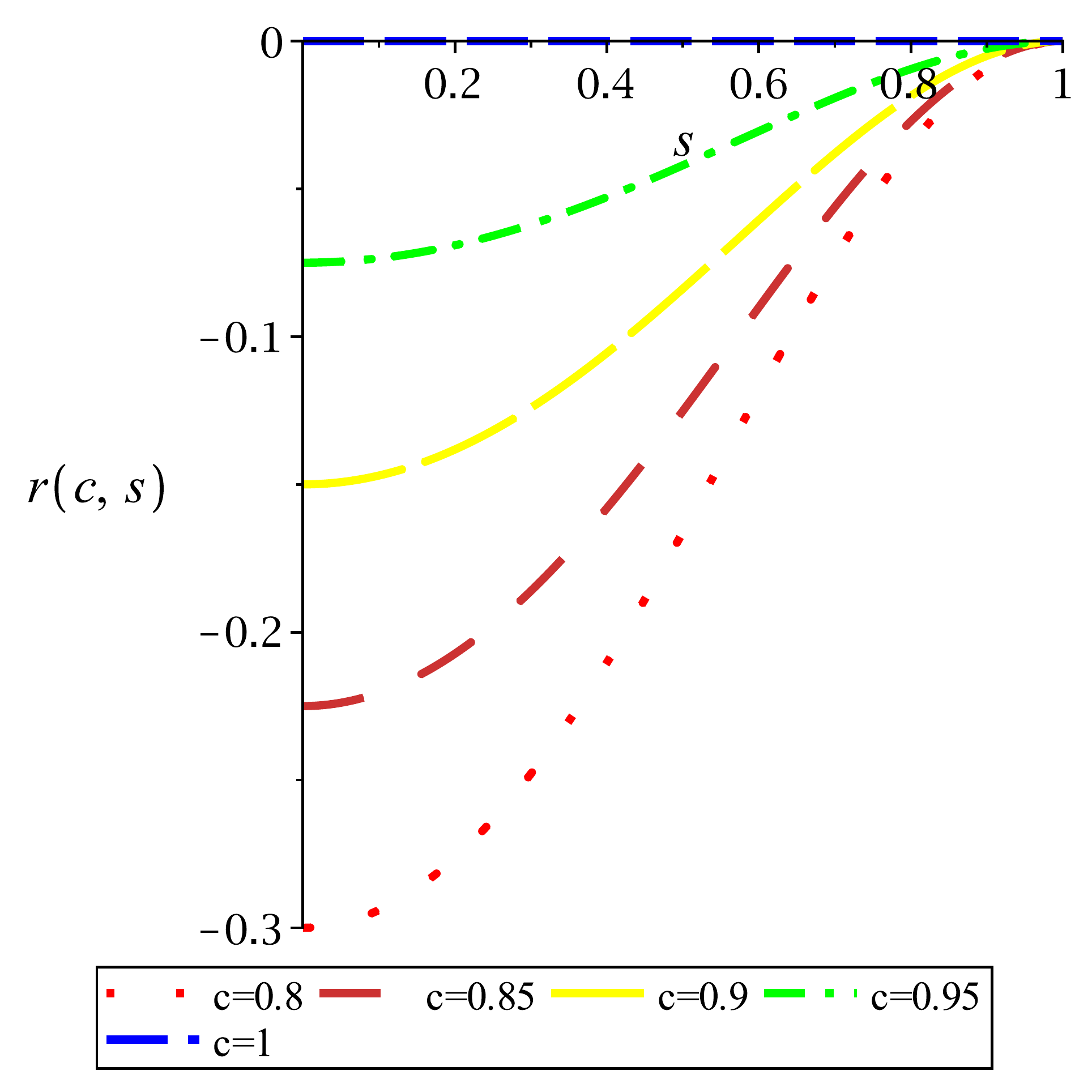}
        \caption{Behaviour of $r(c,s)$, Eq. (\ref{3.7}), for different concentrations $c$ of spins with $L=1$. From bottom up: $c=0.8;\, 0.85;\,
        0.9;\, 0.95;\, 1$ (red, orange, yellow, green and blue curves, coloured online). Value $r(c,s=0)=-\frac{3}{2}(1-c)$ corresponds to the diluted
        Ising model with non-magnetic impurities.
        \label{fig2}}
    \end{center}
\end{figure}

Since the scale of the spin length is fixed by Eq. (\ref{3.4}), one gets an obvious relation:
\begin{equation} \label{3.8}
r(c,s)=r(1-c,1/s)\, .
\end{equation}
Therefore, without a loss of generality we restrict ourselves in further analysis
to the region $0<s<1$. Behavior of $r(c,s)$ for several values of concentration $c$ of `longer' spins with $L=1$ is shown in Fig. \ref{fig2}.

In the next section we will use available knowledge about RG functions of the model with the effective Hamiltonian
(\ref{3.3}) together with obtained here dependencies of its bare couplings on spin concentration and length
to discuss the effective critical behaviour of the random spin length Ising model.

\section{\label{IV}Renormalization group flows and critical exponents}

Renormalization group approach is known to be a standard and powerful tool to calculate  universal
characteristics of critical behaviour, like asymptotic critical exponents and critical amplitude ratios
\cite{Amit,rgbooks}.
However, this approach has also been used for a quantitative description of critical behaviour outside
an asymptotic regime, where the non-universal effective critical exponents (\ref{1.2}) do not yet attain their
universal asymptotic
values (\ref{1.1}). An example is given by a study of flows of couplings under
renormalization that describes  experimentally observed  effective (non-universal) behavior of
the amplitude ratio of thermal conductivity in $^4$He, transport coefficients in $^3$He-$^4$He mixture,
shear viscosity in Xe  (see review \cite{Folk06} for these and other examples).
Similar approach has been used to study effective critical behaviour of diluted uniaxial (Ising-like)
\cite{Janssen95,Folk00,Calabrese04} and Heisenberg \cite{Dudka03,Calabrese04} magnets.
In particular, the field-theoretical RG has been used to explain an experimentally
observed impact of concentration of a non-magnetic component on the phase transition
into magnetically ordered state. Obtained results show that scenario of an effective
critical behaviour depends on the initial couplings of the effective Hamiltonian,
that in their turn are dependent on the disorder strength as well.

Before proceeding further, let us rewrite the effective Hamiltonian (\ref{3.6})
in a standard notation used in field theoretical RG analysis:\footnote{We use notations of Ref. \cite{Kompaniets21}.}
\begin{equation}\label{effham}
{\cal H}_{\rm eff} {=} {-}\!\int\!\! d^d {\bf R}\!
\left\{\!\frac{1}{2}\sum_{\alpha{=}1}^n\left[m_0^2(\phi^\alpha_{\bf R})^2{+}
|\vec{\nabla}\phi^\alpha_{\bf R}|^2\right]\!{+}
\frac{g_{1,0}}{4!}\sum_{\alpha,\beta=1}^n  (\phi^\alpha_{\bf R})^2  (\phi^\beta_{\bf R})^2+
\frac{g_{2,0}}{4!} \sum_{\alpha=1}^n  (\phi^\alpha_{\bf R})^4
\!\right\},
\end{equation}
where continuous limit has been taken, squared unrenormalized mass $m_0^2$
measures the temperature distance to the critical point, and the ratio of
bare couplings $g_{1,0}/g_{2,0}$ equals to  $r(c,s)$, see Eq. \ref{3.7}.
A change in couplings $g_1$, $g_2$
 under the renormalization is described by the flow equations \cite{Amit,rgbooks}:
\begin{equation} \label{flows}
\ell\frac{\rm d}{{\rm d} \ell}g_1(\ell)=\beta_{g_1}\left(g_1(\ell),g_2(\ell)\right),\quad
\ell\frac{\rm d}{{\rm d}\ell}g_2(\ell)=\beta_{g_2}\left(g_1(\ell),g_2(\ell)\right),
\end{equation}
where $\ell$ is the flow parameter. The flow parameter can be related by means of a so-called `matching
condition' to the correlation length, depending on temperature distance $\tau$ to the critical
point, see e.g., Ref. \cite{Folk06}. In simpler case it is proportional to the inverse squared correlation length, therefore the limit
$\tau\to 0$ corresponds to $\ell\to 0$. The fixed points ($g_1^*,g_2^*$) of the system of differential equations
(\ref{flows}) are given by:
\begin{equation} \label{6}
\beta_{g_1}\left(g_1^*,g_2^*\right)=0,\quad
\beta_{g_2}\left(g_1^*,g_2^*\right)=0.
\end{equation}
A fixed point (FP) is said to be stable if the stability matrix
\begin{equation}\label{6a}
B_{ij}\equiv\partial \beta_{g_i}/\partial g_j, \hspace{3em}
i,j=1,2,
\end{equation}
possesses in this point eigenvalues $\omega_1,\omega_2$ with positive real parts.
In the limit $\ell\to 0$, $g_1(\ell)$ and $g_2(\ell)$ attain the
stable FP values $(g_1^*,g_2^*)$. If the stable FP
is reachable from the initial conditions (let us recall that for the
effective Hamiltonian (\ref{effham}) the initial conditions depend on
$c$ and $s$ and lie in the region $g_2> 0, g_1\leq 0$) it corresponds to
the critical point of the system. The asymptotic critical exponents
are defined by the FP values of the RG $\gamma$-functions. In particular, the
isothermal magnetic susceptibility and correlation length exponents $\gamma$
and $\nu$ are expressed in terms of the
RG functions $\gamma_{\phi}$ and ${\gamma}_{m^2}$ describing renormalization of the field $\phi$
and of the  squared mass $m^2$
correspondingly \cite{Amit,rgbooks}:
\begin{equation} \label{gam_as}
\gamma=\frac{2-2{\gamma}^*_{\phi}}{2+{ \gamma}^*_{m^2}},\qquad\nu=\frac{1}{2+{ \gamma}^*_{m^2}}.
\end{equation}
In Eq. (\ref{gam_as}), the superscript means that the corresponding $\gamma$-functions are calculated at
a stable FP: $\gamma^*_{\phi}\equiv \gamma_{\phi}(g_1^*,g_2^*)$,
${ \gamma}^*_{m^2}\equiv{
\gamma}_{m^2}(g^*_1,g^*_2)$. In the RG scheme, the effective critical exponents
are calculated in the region,
where the couplings $(g_1(\ell),g_2(\ell))$ have not yet reached their fixed point values
and depend on $\ell$. In particular, for the exponents $\gamma_{\rm eff}$ and $\nu_{\rm eff}$ one
gets \cite{Folk00}:
\begin{equation}\label{gam}
\gamma_{\rm eff}(\tau)= \frac{2-2{\gamma}_{\phi}
[g_1\{\ell(\tau)\},g_2\{\ell(\tau)\}]}{2+{ \gamma}_{m^2}
[g_1\{\ell(\tau)\},g_2\{\ell(\tau)\}]}+\dots,\quad\nu_{\rm eff}(\tau)= \frac{1}{2+{ \gamma}_{m^2}
[g_1\{\ell(\tau)\},g_2\{\ell(\tau)\}]}+\dots\, .
\end{equation}
In (\ref{gam}) the  part denoted by dots
is proportional to the $\beta$-functions
 and comes from the change of the amplitude parts of susceptibility and of
 correlation length.
In the subsequent calculations we will neglect these parts, taking
the contribution of the amplitude function to the crossover to be small, as it was used in other studies of disordered systems
\cite{Janssen95,Folk00,Dudka03}. However,  a more sophisticated approach was developed in Ref.~\cite{Calabrese04} to
quantitatively check dependence of effective critical exponents directly on temperature distance to  the critical point.

In our study we use the most up-to-date  RG functions obtained for the effective Hamiltonian (\ref{effham}) with a record six-loop accuracy in
the minimal subtraction renormalization scheme \cite{Kompaniets21}  (actually,  as a partial case of the RG functions for the
cubic model \cite{Adzhemyan19}). The series read:
\begin{eqnarray}\label{b1}
\beta_{g_1}(g_1,g_2)&=&-{g_1} \left(\varepsilon-\frac{8 {g_1}}{3}-2 {g_2}+\frac{14 g_1^2}{3}+\frac{22 {g_1} {g_2}}{3}+\frac{5 g_2^2}{3} +\dots\right),
\\
\beta_{g_2}(g_1,g_2)&=&-{g_2} \left(\varepsilon-4 {g_1}-3 {g_2} +\frac{82 g_1^2}{9}+\frac{46 {g_1} {g_2}}{3}+\frac{17 g_2^2}{3}+\dots\right),\label{b2}
\\
\gamma_{\phi}(g_1,g_2)&=&\frac{g_1^2}{18}+\frac{{g_1} {g_2}}{6}+\frac{g_2^2}{12}+\dots\label{gam1}
\\  \label{gam2}
\gamma_{m^2}(g_1,g_2)&=&- g_2-\frac{2 g_1}{3}+\frac{5 g_1^2}{9}+\frac{5 {g_1} {g_2}}{3}+\frac{5 g_2^2}{6}+\dots
\end{eqnarray}
Here, $\varepsilon=4-d$ and the dots indicate higher order terms currently known up to $O(g_i^7)$ and $O(g_i^6)$ for
the $\beta$- and $\gamma$-functions correspondingly \cite{Kompaniets21,Adzhemyan19}.

\begin{figure}[ht]
\begin{center}
\includegraphics[width=0.6\paperwidth]{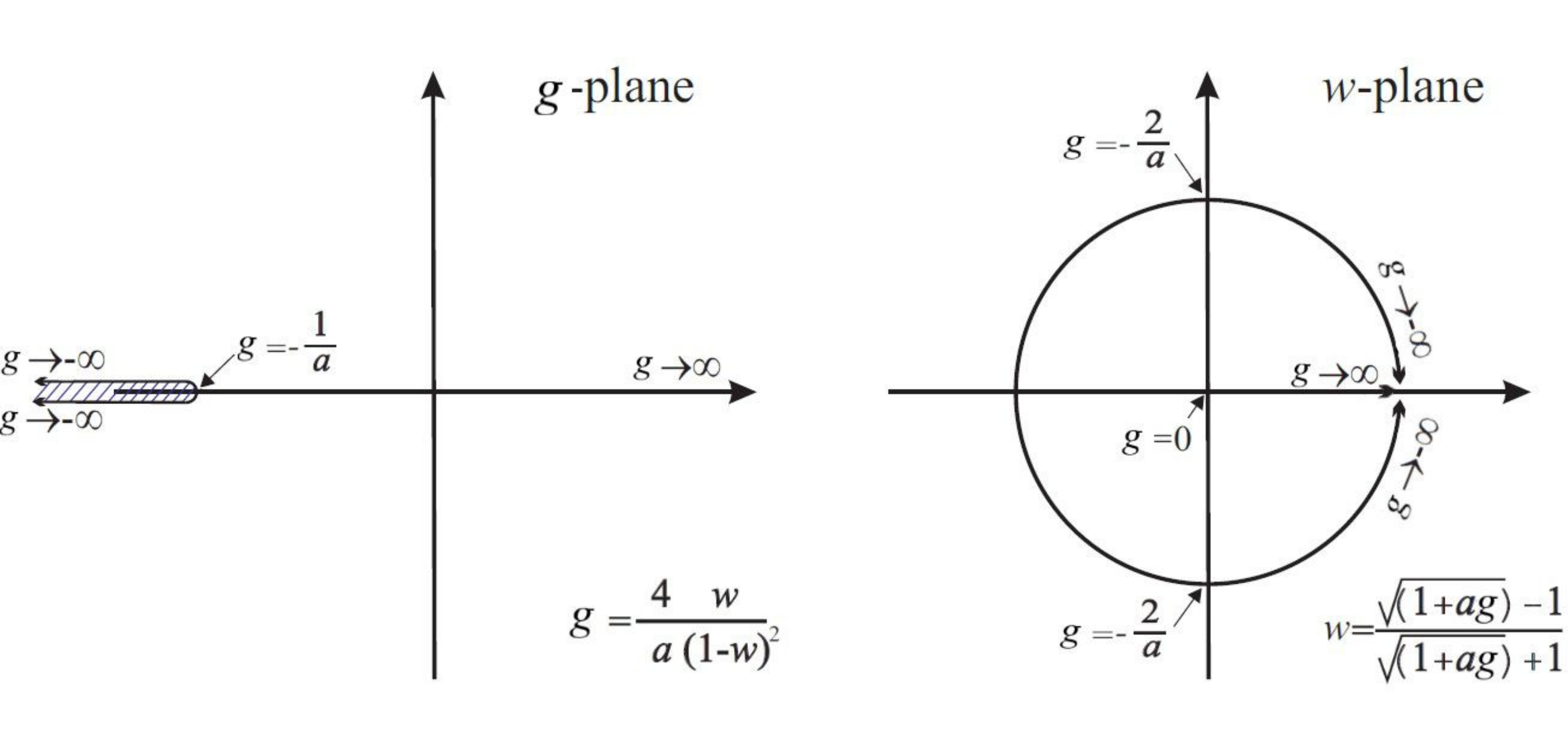}
\end{center}
\caption{ Conformal mapping of the cut-plane onto a disc. See the
text for details. \label{confor3} }
\end{figure}

Starting form the RG expressions  one can
either develop the $\varepsilon$-expansion, or work directly at
$d=3$ considering the renormalized couplings $(g_1,g_2)$ as the expansion
parameters \cite{Schloms1,Schloms2}.
However, such RG perturbation theory series
are known to be asymptotic at best \cite{rgbooks}.
One should apply appropriate resummation technique to improve their convergence to get
reliable numerical data on their basis. Many different resummation procedures are currently
 used to this end, see e.g., \cite{Folk03,Pelissetto02}.
In our study we use the method, based on the Borel
transformation combined with a conformal mapping \cite{LeGuillou77}.  This resummation
technique was first elaborated for field-theoretical models with one coupling. It
requires knowledge about a high-order behaviour of the series.
In particular, for the series
\begin{equation}
f(g)=\sum_{l=0}^\infty a_l  g^l  \label{series1}
\end{equation}
with the known large-order behaviour of the coefficients
\begin{equation}\label{grow}
a_l=(-a)^l  l^b l!~ \left[ 1+O(1/l)\right], \quad l\to\infty ,
\end{equation}
where numbers $a$ and $b$ characterize the main divergent part,
one can build its resummed counterpart as:
\begin{equation}
f_{\mathrm{R}}(g)=\sum_{l=0}^\infty d_l(a, b) \int_0^{\infty}
d t \,\,{{\rm e}^{-t}\, t^{b}\ \left[w(g
t)\right]^l}\ . \label{resummation1}
\end{equation}
In Eq. (\ref{resummation1})     $d_l(a,b)$   are the coefficients of the re-expansion of
the Borel-Leroy image of $f(g)$:
\begin{equation} \label{resummation1b}
B(g)=\sum_{l=0}^\infty \frac{a_l}{\Gamma(l+b+1)} \ g^l\,
\end{equation}
which can be written in powers of the new variable $w(g)={\sqrt{1 + a\, g}-1\over \sqrt{1 + a\, g}+1}$:
\begin{equation} \label{resummation1c}
B(\omega)=\sum_{l=0}^\infty  d_l(a,b) \omega^l\, .
\end{equation}
$\Gamma(x)$ in (\ref{resummation1b}) is Euler gamma function.
This transformation is a conformal  mapping of the  complex
$g$-plane cut from  $g=-1/a$  to $-\infty$
onto the unit circle in the $w$-plane such that the singularities
of $B(g)$ lying  on the negative axis   now lie on   the boundary
of the  circle $|w|=1$, see Fig. \ref{confor3}.

To impose the strong coupling behavior of the series, $f(g\to \infty)\sim g^{\alpha/2}$, Eq. (\ref{resummation1}) needs to be
generalized in the
following way  \cite{kazakov79}
\begin{equation}
f_{\mathrm{R}}(g)=\sum_{l=0}^\infty d_l(\alpha,a, b)  \int_0^{\infty}
d t\,\, {{\rm e}^{-t}\,  t^{b}}\ { \left[w(g t)\right]^l \over \left[1-w(g t)\right]^{\alpha} }\, .
\label{resummation2}
\end{equation}

This procedure can be extended  for  field-theoretical
models that contain several couplings. For instance, when $f$ is a function  of two
variables  $g_1$ and $g_2$,  the  resummation  technique can treat $f$ as a
function of coupling associated with terms related to the pure (undisordered) magnet (in our case it is $g_2$) and
a ratio $z=g_1/g_2$, \cite{Alvarez00,Pelissetto00,Pelissetto04}:
\begin{equation}
f(g_2,z)=\sum_{l=0}^\infty a_l(z) \ g_2^l \, . \label{series}
\end{equation}
 Then, keeping $z$ fixed and performing the
resummation only in $g_2$   according to the steps described above, one gets:
\begin{equation}
f_{\mathrm{R}}(g_2,z)=\sum_{l=0}^\infty d_l(\alpha,a(z),b;z)
\int_0^{\infty} d t\,\,{{\rm e}^{-t}\, t^{b}}{
\left[w(g_2 t;z)\right]^n \over \left[1-w(g_2
t;z)\right]^{\alpha} } \, , \label{resummation}
\end{equation}
with $z$-dependent parameter $a(z)$. Here,  as     above,  the  coefficients
$d_l(\alpha,a(z),b,z)$ in (\ref{resummation})  are computed  so
that   the  re-expansion of  the right hand side  of (\ref{resummation}) in
powers of $g_2$ coincides with that of (\ref{series}).

Calculation of  parameters $a, b, \alpha$ for the field-theoretical models with several couplings
is a complicated  task. Furthermore, in practice,  procedure described above is applied to the
truncated series (\ref{b1})-(\ref{gam2}), which are known up to the corresponding order $L$.  As a result,
the resummed expansions (that usually correspond to certain physical observable) appear to be dependent
on resummation parameters $a$, $b$ and $\alpha$.
Often applying the conformal Borel method for models with two couplings, such
parameters  are taken in the region where physical observables are less sensitive to their values,
see e.g., Ref. \cite{Delamotte08}.
Although Borel summability is not proven for perturbative series  for models with
disorder in $d>0$ \cite{Alvarez00}, the conformal Borel method has been successfully used
for random Ising model \cite{Pelissetto00,Pelissetto04,Folk06,Krinitsyn06} providing accurate critical exponents  in agreement with previous analytical estimates.
In our analysis we test the region of parameters discussed  in the study of five-loop series for the model
of frustrated magnet  \cite{Delamotte08}: $a=1/2$, $b$ is
varying in the interval  $[6, 30]$, while $\alpha$  is varying in $[-0.5, 2]$.
In what follows below we will give more precise values of the parameters.

\begin{figure}[h]
    \begin{center}
        \includegraphics[width=0.35\paperwidth]{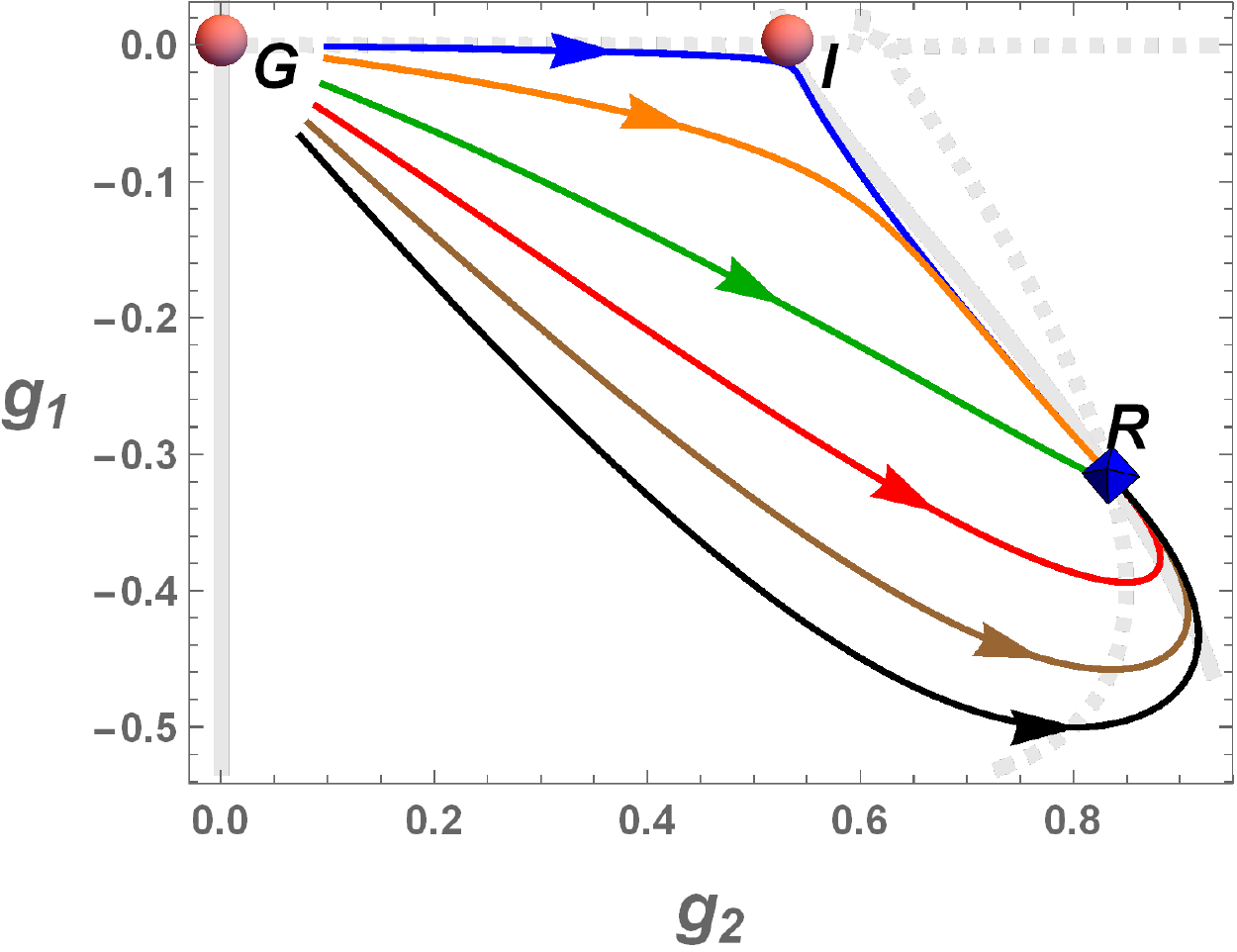}
        \caption{Renormalization group flow in the space of couplings for different initial conditions. From up down: $r=g_1(\ell_0)/g_2(\ell_0)$= -0.01 (blue), -0.1 (orange), -0.3 (green), -0.5 (red), -0.7 (brown), -0.9 (black). Unstable Gaussian (G) and Ising (I) fixed points (shown by red discs) as well as a stable random Ising (R) fixed point  (shown by the blue diamond) are located at the intersection points of lines of zeros for functions $\beta_{g_1}$ (dashed shadow lines) and $\beta_{g_2}$ (solid shadow lines) \label{floww}}
    \end{center}
\end{figure}

To analyze effective critical behaviour, we apply the above described conformal Borel resummation to the
$\beta$-functions (\ref{b1}), (\ref{b2}) in the six-loop approximation. In the region of couplings we are interested in
one finds three FPs as solutions of Eqs. (\ref{6}): Gaussian FP {\bf G} ($g_1^*=g_2^*=0$), pure Ising FP {\bf I}
($g_1^*=0,\,g_2^*\neq 0$), and random Ising FP {\bf R} ($g_1^*\neq 0,\, g_2^*\neq 0$), as shown in Fig.
\ref{floww}.
For the cases when random Ising FP exists and is a stable locus, our results  are compatible with results
of Ref. \cite{Kompaniets21} $(g_1^*\sim -0.3 , g_2^*\sim 0.8)$. As a final choice of the resummation parameters
we get $a=1/2$, $b=10$ and $\alpha=1$. We will see below that at such parameter values the asymptotic
critical exponents of the random Ising model are very close to the latest analytical estimates
(see the Table~\ref{tab1} and the review in Ref. \cite{Kompaniets21}): $\gamma=1.333$ and $\nu=0.676$.
Solving numerically the system of differential equations (\ref{flows}) we get the running values of the couplings
$(g_1(\ell),g_2(\ell))$. They define the flow in the parametric space $g_1$, $g_2$ and in
the limit $\ell \to 0$ attain the stable fixed point value (shown by the blue diamond
in Fig.~\ref{floww}).

The properties of the flow depend on the initial conditions $(g_1(\ell_0)$, $g_2(\ell_0))$  for
solving the system of differential equations (\ref{flows}).
Typical flows obtained for different
ratios $g_1(\ell_0)/g_2(\ell_0)=r$ are shown in Fig.~\ref{floww} by curves of different color. We choose the starting values
in the region with the appropriate signs of couplings $g_1<0$, $g_2>0$ near the
origin (in the vicinity of the Gaussian fixed point {\bf G} shown by the
red disc in the Fig.~\ref{floww}).
As one sees from the figure, although all RG flows with the decrease of $\ell$ lead to the stable FP, depending on the
initial conditions they manifest very different behaviour. For small values of $r$ the flows may first stay within the
basin of attraction  of the unstable Ising FP (blue curve), with an icrease of $r$ they directly approach
the stable random FP (green curve), and further increase of $r$ leads to the `overshooting' behaviour, when the running
values of the couplings approach their stable FP values from above (black, brown curves).
Let us further analyse, how the observed behaviour of the RG flows is
manifest in the effective critical behaviour, in particular, in the values
of effective critical exponents.

\begin{figure} [h]
    \begin{center}
        \includegraphics[width=0.35\paperwidth]{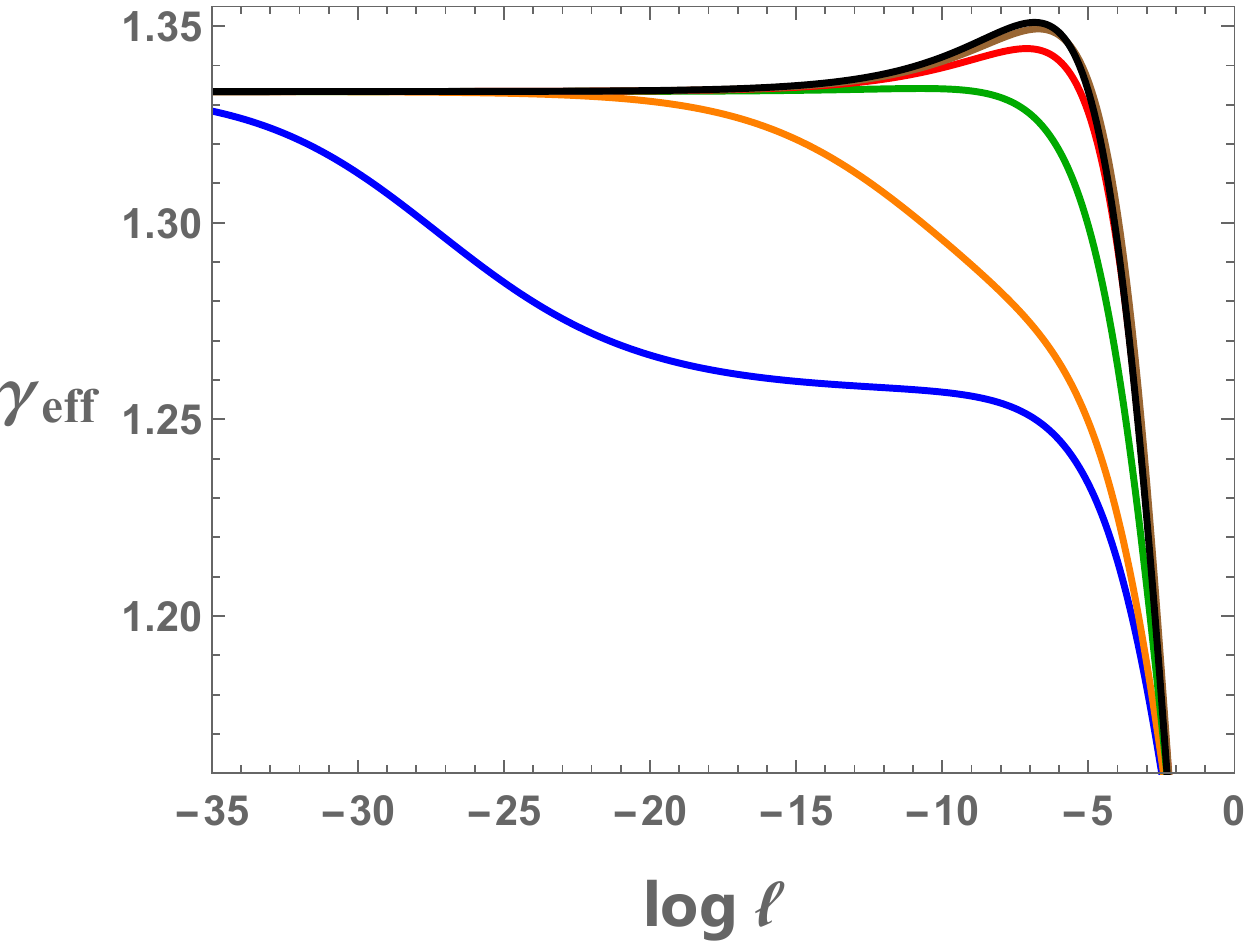}\hspace{3em}
         \includegraphics[width=0.35\paperwidth]{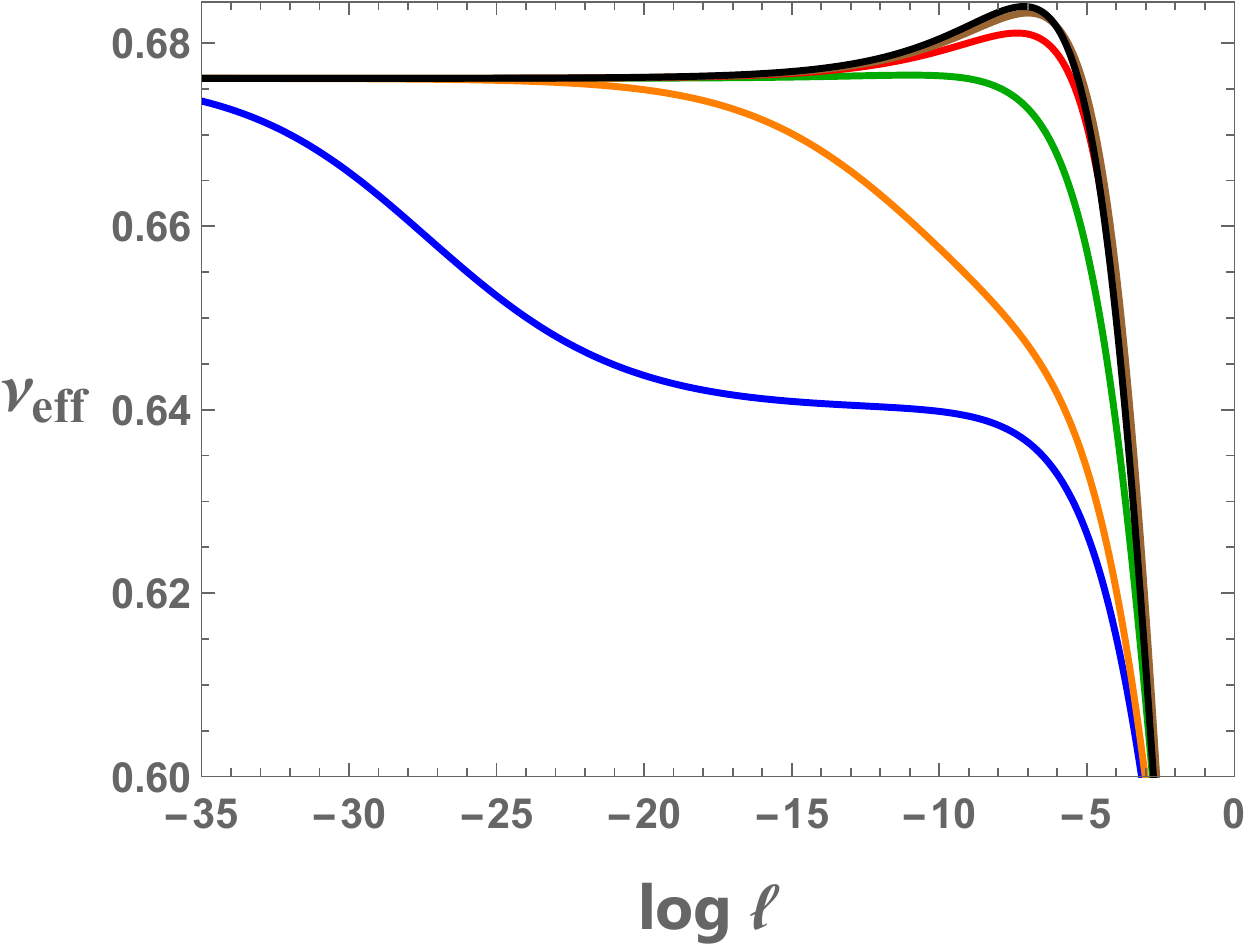}
         \\ \vspace{3ex}
\centerline{\bf (a) \hspace{20em} (b)} \vspace{3ex}
        \caption{Dependencies of   the magnetic susceptibility effective critical exponent $\gamma_{\rm eff}$ ({\bf a}) and the
        correlation length effective critical exponent $\nu_{\rm eff}$ ({\bf b}) on the flow parameter
        calculated along the RG flows of Fig. \ref{fig2}. \label{exponn}}
    \end{center}
\end{figure}

Indeed, the running values of couplings presented by flows in Fig.~\ref{floww} allow
to get access to the effective critical exponents. To this end, we apply the above described resummation procedure to the six-loop expansions for the  effective critical exponent  (\ref{gam}) with the flow-parameter dependent couplings. In
this way we get dependencies of the
isothermal susceptibility and correlation length effective critical exponents on the flow parameter $\ell$. These are shown in Fig.~\ref{exponn} by corresponding colors.
Different regimes in behaviour of couplings $(g_1,\, g_2)$  in the course of reaching the stable FP are reflected in different regimes
observed for the effective critical exponents when the critical temperature is reached. Depending on the initial conditions,
the effective critical exponents either reach their universal values at the stable FP comparatively fast  (green curves) or attain the values
that differ from the stable FP ones in a broad crossover region. In turn, the crossover region may be either governed
by the pure Ising model universality class critical exponents (blue curve)
or by the values of the exponents that exceed the stable FP ones
(peaks in black and red curves),

For our model (\ref{effham}), the ratio of the initial  couplings
depends on concentration $c$ and spin length $s$, $r=r(c,s)$ via Eq. (\ref{3.7}).
Therefore, changing $c$ and $s$ one is able to discriminate between different regimes
in the behaviour of the effective critical exponents. To give an example,
taking the six-loops stable FP values for $g_i^*$ from Ref. \cite{Kompaniets21} obtained
for different resummations parameters one gets the following estimate
for the region of $r$ at the FP: $-0.4 < r ^* < -0.25$. On the other hand,
from Eq. (\ref{3.7}) one gets the following result for bare (initial) $r$
of the diluted Ising model with concentration of magnetic sites $c=0.8$:  $r(c=0.8, s=0)=-0.3$.
The closeness of this value  to its renormalized counterpart may suggest a direct flow of
couplings under renormalization to the stable FP, cf. the green curve in Fig. \ref{floww}.
Such behaviour without almost no correction-to-scaling terms for $c=0.8$ has indeed been
observed for this model in MC simulations  \cite{Ballesteros98,Calabrese03}.

\section{\label{V}Conclusions and outlook}

In this article, we set ourselves a twofold task. On the one hand, our goal was to
show how the disorder in structure alone, without presence of a non-magnetic component
can modify the universality class of paramagnetic-ferromagnetic phase transition
in uniaxial magnets. On the other hand, we wanted to analyze possible changes
in the non-universal effective critical behaviour in such systems. To this end,
we considered a random spin length Ising model, when the length $L$ of each spin
is a random  variable  governed by the distribution function $p(L)$
\cite{Krasnytska20,Krasnytska21}.

It is instructive to note, that this model has originally been suggested in the
context of sociophysics, where its intention was to describe social interactions
and opinion formation between agents of binary nature but of different opinion strength.
Here, we used this model to explain  phenomenon within the condensed matter physics.
We showed that particular cases of this model,
a site-diluted Ising model and an Ising model with spins of two different lengths,
are described by the effective Hamiltonians (\ref{3.3}) and (\ref{3.6}) of the same symmetry, hence they belong
to the same universality class. This means, that their asymptotic critical exponents
(\ref{1.1}) are the same.
However, effective critical behaviour of these systems differs.
We made use of the field-theoretic RG approach to describe such difference
quantitatively. Doing so, we reproduce with a high precision the known values
of the asymptotic critical exponents in the universality class of the 3d diluted Ising
model (compare our results $\gamma=1.333$ and  $\nu=0.676$ with that of Ref.
\cite{Pelissetto00} given in the fourth line of Table \ref{tab1}).

Furthermore, beyond the narrow region in the
vicinity of $T_c$, one may observe very different effective critical behaviour even within
the same universality class, as we explicitly show in Fig. \ref{exponn}.  The effective critical
exponents shown there change as $T_c$ is approached and, strictly speaking, attain
their universal values only at $T_c$. Changes in the values of the effective exponents, together with
distance to $T_c$ depend also on the initial conditions imposed when solving differential
equations for the RG flows. The latter, in turn, can be attributed to the microscopic features
of the magnet, given in our case by the concentration $c$ and ratio of spin lengths $s$. These
features are encoded in the ratio (\ref{3.7}) that governs initial conditions for the RG flows. Thus, the
scheme implemented here enables one to analyse effective critical behaviour of structurally
disordered magnets in a self-consistent way.

The effective critical exponents discussed in this paper can
also appear when the critical behaviour is analysed via numerical simulations.
Indeed, consider the quotient method~\cite{XIL,Extra,Amit} as an example. In this method one computes the  temperature, $T_c(L)$, at which a
dimensionless observable, $g(L,T)$ (for instance the Binder cumulant, $B(L,T)$ or the correlation length in units of the lattice size
$\xi(L,T)/L$) crosses for lattice sizes $L$ and $2L$: thus, $g(L,T_c(L))=g(2L, T_c(L))$. Moreover, one can  also consider a dimensionful
observable which diverges at the critical point as ${\cal O}(L,T_c)\sim L^{x/\nu}$, where $\nu$ is the correlation length critical exponent. By computing the ratio ${\cal O}(2L,T_c(L))/{\cal O}(L,T_c(L))$ is possible to write
\begin{equation}
\label{eq:quotient}
    \left.\frac{x}{\nu}\right|_{\infty}=\left.\frac{x}{\nu}\right|_{(L,2L)}+ O( \frac{1}{L^\omega})\, ,
\end{equation}
where $\omega$ can be identified with the leading irrelevant eigenvalue in the RG approach~\cite{Amit,rgbooks} and
\begin{equation}
\left.\frac{x}{\nu}\right|_{(L,2L)} \equiv \log_2 \frac{{\cal O}(2L,T_c(L))}{{\cal O}(L,T_c(L)}\,,
\end{equation}
which can be used as the definition of the effective critical exponent.
Notice that in general, in Eq. (\ref{eq:quotient}) will not only appear the leading irrelevant eigenvalue of the RG, $\omega$, but all the irrelevant eigenvalues, so the behavior of the effective exponent will not be monotonic in most of the cases. An example of this non-monotonic behavior is the three-dimensional random field Ising model for some observables and different choices of the random magnetic field~\cite{Fytas}.
Finally, we remark that the finite size scaling theory can be understood using a RG transformation with a RG parameter, $\ell$, given by  $\ell=1/L$~\cite{Amit}.

A natural continuation of our study will be the
numerical simulation of the random Ising model with two different spin lengths, section \ref{IIb}
and analysis of its asymptotic and effective critical behaviour. And -- last but not least -- search
for experimental realizations of the phenomena discussed in this paper.

\section*{Acknowledgements}
We thank participants of an annual Atelier ``Statistical Physics and Low-dimensional Physics''
in Pont-\`a-Mousson and of the seminar of Compexity Science Hub in Vienna where these results
were presented. We acknowledge useful discussions with  Bertrand Berche, Reinhard Folk, Ralph Kenna,
Mykola Shpot, and Stefan Thurner.
This work was partially supported by  National Academy of Sciences of Ukraine, project
K$\Pi$KBK 6541030.
MK acknowledges support of the PAUSE programme and  the hospitality of the Laboratoire de Physique et Chimie Th\'eoriques, Universit\'e de Lorraine
(France) and of the University of Extremadura (Spain).
JJRL acknowledges support by Ministerio de Ciencia, Innovaci\'on y Universidades (Spain), Agencia Estatal de Investigaci\'on (AEI, Spain), and Fondo Europeo de Desarrollo Regional (FEDER, EU) through the Grant PID2020-112936GB-I00 and by the Junta de Extremadura (Spain) and Fondo Europeo de Desarrollo Regional (FEDER, EU) through Grants No.\ GR21014 and IB20079.  YuH acknowledges support of the JESH mobility program of the
Austrian Academy of Sciences and hospitality of the Complexity Science Hub Vienna when finalizing this paper.

\end{document}